\documentclass[12pt]{article}
\usepackage{color}
\usepackage{latexsym}
\usepackage{epsfig,amssymb,euscript, mathrsfs}
\usepackage{amsmath}
\newsavebox{\ns}
\newsavebox{\dbrane}
\newsavebox{\dbshort}

\def\be{\begin{equation}}
\def\ee{\end{equation}}
\def\bea{\begin{eqnarray}}
\def\eea{\end{eqnarray}}

\newcommand{\nn}{\nonumber}

\newcommand\R{\mathbb{R}}
\newcommand\Z{\mathbb{Z}}

\newcommand\C{\mathbb{C}}

\newcommand\diff{\mathrm{d}}
\newcommand{\de}{\partial}

\newcommand{\dd}{\mathrm{d}}
\newcommand{\me}{\mathrm{e}}
\newcommand{\ii}{\mathrm{i}}

\newcommand{\ex}{\mathrm{e}}
\newcommand{\vol}{\mathrm{vol}}

\newcommand{\Imag}{\mathrm{Im}\, }
\newcommand{\Real}{\mathrm{Re}\, }

\newcommand{\proj}{{\cal P}_-}

\numberwithin{equation}{section}       

\begin{document}

\begin{titlepage}

\begin{center}

\today

\vskip 2.3 cm 

{\Large \bf The free energy of $\mathcal{N}=2$ supersymmetric}
\vskip .5cm 
{\Large \bf AdS$_4$ solutions of M-theory}

\vskip 2 cm

{Maxime Gabella$^1$, Dario Martelli$^2$, Achilleas Passias$^2$ and James Sparks$^3$\\}

\vskip 1cm

$^1$\textit{Rudolf Peierls Centre for Theoretical Physics,
University of Oxford, \\
1 Keble Road, Oxford OX1 3NP, United Kingdom\\}

\vskip 0.8cm

$^2$\textit{Department of Mathematics, King's College, London, \\
The Strand, London WC2R 2LS,  United Kingdom\\}

\vskip 0.8cm

$^3$\textit{Mathematical Institute, University of Oxford,\\
24-29 St Giles', Oxford OX1 3LB, United Kingdom\\}

\end{center}

\vskip 2 cm

\begin{abstract}
\noindent We show that general $\mathcal{N}=2$ supersymmetric AdS$_4$ solutions of 
M-theory with non-zero M2-brane charge admit a canonical contact structure. 
The free energy of the dual superconformal field theory on $S^3$ and the 
scaling dimensions of operators dual to supersymmetric wrapped M5-branes 
are expressed via AdS/CFT in terms of contact volumes. In particular, this leads
to topological and localization formulae for the coefficient of $N^{3/2}$ in 
the free energy of such solutions.

\end{abstract}

\end{titlepage}

\pagestyle{plain}
\setcounter{page}{1}
\newcounter{bean}
\baselineskip18pt

\section{Introduction}

Tremendous progress has been achieved recently in understanding the AdS$_4$/CFT$_3$ correspondence, following the important results 
of \cite{Bagger:2006sk,Gustavsson:2007vu,Aharony:2008ug}.
In particular, for ${\cal N}\geq 2$ supersymmetry there is often good control on both sides of the correspondence.
On the gravity side, the simplest setup
is that of Freund-Rubin AdS$_4\times Y_7^\text{SE}$  backgrounds of M-theory where $Y_7^\text{SE}$ is a Sasaki-Einstein manifold\footnote{Particular cases with ${\cal N}>2$ include
3-Sasakian manifolds and orbifolds of the round seven-sphere.}, and deformations thereof. These are conjectured to be dual to the theory on a 
large number of multiple M2-branes
placed at a Calabi-Yau four-fold singularity. Rather generally,  these field theories are believed to be strongly coupled Chern-Simons-matter theories 
at a conformal fixed point. 

While gravity computations are relatively amenable, obtaining results directly in the three-dimensional strongly coupled field theories 
has been prohibitively difficult until very recently. For this reason, non-trivial quantitative tests of the AdS$_4$/CFT$_3$ correspondence were not available.
The situation has improved considerably with the results of \cite{Jafferis:2010un,Hama:2010av}
(based on \cite{Kapustin:2009kz}), who showed that the partition function $Z$ of ${\cal N}=2$ supersymmetric field theories on the three-sphere 
can be reduced to more manageable \emph{matrix integrals} using \emph{localization} techniques. Moreover, in  \cite{Jafferis:2010un} it has been conjectured
that at a conformal fixed point the \emph{free energy}, defined as ${\cal F}= -\log |Z|$, is extremized as a function of all possible R-symmetries. 
If this is true, this quantity would
then be analogous to the central charge $a$ of four dimensional SCFTs. More generally, there are expectations that the free energy is a good measure 
of the number of degrees of freedom of three-dimensional field theories, even without supersymmetry.  

In \cite{Martelli:2011qj, Cheon:2011vi, Jafferis:2011zi} the leading large $N$ contribution to the
free energy of Chern-Simons-matter theories on $S^3$ was computed for large classes of ${\cal N}=2$  theories, and
succesfully matched to the gravity prediction in a class of Sasaki-Einstein geometries. 
This remarkable matching was first obtained in \cite{Drukker:2010nc}
for the ABJM theory and then in \cite{Herzog:2010hf} for several ${\cal N}=3$ examples.
In this paper we will derive an  expression for the (holographic) free energy ${\cal F}$, 
valid for a very general class of  AdS$_4\times Y_7$ solutions dual to  
${\cal N}=2$ three-dimensional SCFTs. 
In fact, we will consider the most general class of M-theory AdS$_4$ solutions 
with non-zero M2-brane charge, finding very similar results to the type IIB AdS$_5$ geometries 
with non-zero D3-brane charge in \cite{Gabella:2009ni}. We will prove the geometric formula
\bea\label{freeenergy}
\mathcal{F} &=& N^{3/2} \sqrt{\frac{32\pi^6}{9\int_{Y_7} \sigma\wedge (\diff\sigma)^3}}~,
\eea
where $N$ is the quantized M2-brane charge and $\sigma$ is a particular \emph{contact form} on $Y_7$, that we will discuss. 

We will also present a formula for the scaling dimension of BPS operators  $\mathcal{O}_{\Sigma_5}$ dual to 
probe M5-branes wrapped on supersymmetric five-submanifolds $\Sigma_5 \subset Y_7$. 
In particular, the scaling dimension $\Delta(\mathcal{O}_{\Sigma_5})$ of these operators can be calculated from the contact 
volume of the  five-submanifold $\Sigma_5$ as 
\bea
\label{Delta}
\Delta(\mathcal{O}_{\Sigma_5}) &=& \pi N \left|\frac{\int_{\Sigma_5}\sigma\wedge (\diff\sigma)^2}{
\int_{Y_7} \sigma\wedge (\diff\sigma)^3} \right|~.
\eea
Both of these formulae are natural generalizations of those holding in the Sasaki-Einstein case, and are analogous to the results presented in 
\cite{Gabella:2009ni}.

The results of this paper stem from a systematic analysis of the geometry underlying
 general AdS$_4\times Y_7$ M-theory solutions 
preserving at least ${\cal N}=2$ supersymmetry. In particular, we will identify a $\mathtt{u}(1)$  
 symmetry, generated by a  Killing vector field $\xi$, which is 
 the geometric counterpart to the $\mathtt{u}(1)$  R-symmetry of the dual ${\cal N}=2$ superconformal field theory. In addition, 
we will demonstrate the existence of a  \emph{contact structure} on $Y_7$, that will play a key role in deriving  our main results.  
These geometric objects are constructed from the Killing spinors 
preserved by the backgrounds \cite{Gauntlett:2005ww,Gabella:2009ni}, and constitute 
a subset of a canonically defined 
$SU(2)$ structure on $Y_7$. In a subsequent work \cite{WIP} we will present 
the necessary and sufficient 
conditions that this $SU(2)$ structure obeys in order to have an ${\cal N}=2$ supersymmetric solution. 
In \cite{WIP} we will also  present more details of the computations that lead to the results 
discussed here. 
 
\section{Supersymmetric AdS$_4$ solutions of M-theory}

Supersymmetric AdS$_4$ solutions of M-theory have been discussed before 
\cite{Kaste:2003zd,Lukas:2004ip,Behrndt:2005im}; however, we will derive our results without recourse
to the literature.
In \cite{WIP} we will present an analysis of the most general conditions 
for such solutions, in particular focusing on solutions preserving at least ${\cal N}=2$ supersymmetry.  
In this section we summarize 
the Killing spinor equations that are used to derive many of the results
presented  in the remainder of the paper. We refer the reader to \cite{WIP} for further details.

The bosonic fields of eleven-dimensional supergravity consist of  a metric $g_{11}$ and a three-form potential $C$ with 
four-form field strength $G=\diff C$. The signature of the metric is $(-,+,+,\ldots,+)$ and the action is
\bea\label{action}
S &=& \frac{1}{2\kappa^2}\int  R *_{11} \mathbf{1} - \frac{1}{2}G\wedge *_{11}G - \frac{1}{6}C\wedge G \wedge G~,
\eea
where $2\kappa^2 = (2\pi)^8 \ell_p^9$ with $\ell_p$ the eleven-dimensional Planck length.
We consider AdS$_4$ solutions of M-theory of the warped product form
\bea\label{ansatz}
g_{11} &=& \ex^{2\Delta}\left(g_{\mathrm{AdS}_4}+g_{Y_7}\right)~,\nn\\
G&=& m\vol_4 + F~.
\eea
Here $\vol_4$ denotes the Riemannian volume form on AdS$_4$, and without loss of 
generality\footnote{The factor here is chosen to coincide with standard conventions in the case that $Y_7$ is a Sasaki-Einstein seven-manifold. For example, 
the AdS$_4$ metric in global coordinates then reads $g_{\mathrm{AdS}_4} =\tfrac{1}{4}( -\cosh^2\rho\,  \dd \tau^2 + \dd \rho^2 + \sinh^2 \rho\, \dd \Omega^2_2)$.} we 
take $\mathrm{Ric}_{\mathrm{AdS}_4}=-12 g_{\mathrm{AdS}_4}$. In order to preserve the $SO(3,2)$ invariance 
of AdS$_4$ we take $\Delta$ to be a function on the compact seven-manifold $Y_7$.
$F$ is the pull-back of a four-form on $Y_7$, and the Bianchi identity $\diff G=0$ requires that $m$ is
constant and $F$ is closed.

In an orthonormal frame, the Clifford algebra $\mathrm{Cliff}(10,1)$ is generated by gamma matrices $\Gamma_A$ satisfying 
$\{\Gamma_A,\Gamma_B\}=2\eta_{AB}$, where
$A=0,\ldots, 10$, and 
$\eta=\mathrm{diag}(-1,1,\ldots,1)$, and we choose a representation with $\Gamma_0\cdots \Gamma_{10}=1$. The Killing spinor equation is
\bea\label{Killingspinoreqn}
\nabla_M\epsilon + \frac{1}{288}\left(\Gamma_M^{\ \ NPQR}-8\delta_M^N\Gamma^{PQR}\right)G_{NPQR}\, \epsilon &=&0~,
\eea
where $\epsilon$ is a Majorana spinor and $M, N, \ldots$ are spacetime indices. We may decompose $\mathrm{Cliff}(10,1)\cong \mathrm{Cliff}(3,1)\otimes 
\mathrm{Cliff}(7,0)$ via
\bea
\Gamma_\alpha &=& \rho_\alpha\otimes 1~, \qquad \Gamma_{a+3} \ = \ \rho_5\otimes \gamma_a~,
\eea
where $\alpha, \beta= 0,1,2,3$ and $a,b=1,\ldots,7$ are orthonormal frame indices for AdS$_4$ and $Y_7$ respectively, 
$\{\rho_\alpha,\rho_\beta\}=2\eta_{\alpha\beta}$, $\{\gamma_a,\gamma_b\}=2\delta_{ab}$,
 and we have defined $\rho_5=\ii \rho_0\rho_1\rho_2\rho_3$. 
Notice that our eleven-dimensional conventions imply that $\gamma_1\cdots\gamma_7=\ii\, 1$.

The spinor ansatz preserving $\mathcal{N}=1$ supersymmetry in AdS$_4$ is correspondingly
\bea\label{spinoransatz}
\epsilon &=& \psi^+\otimes \ex^{\Delta/2}\chi + (\psi^+)^c\otimes \ex^{\Delta/2}\chi^c~,
\eea
where $\psi^+$ is a positive chirality Killing spinor on AdS$_4$, so $\rho_5\psi^+=\psi^+$, satisfying
\bea
\nabla_\mu \psi^+ &=& \rho_\mu (\psi^+)^c~.
\eea
The superscript $c$ in (\ref{spinoransatz}) 
denotes charge conjugation in the relevant dimension, and 
the factor of $\ex^{\Delta/2}$  is included for later convenience. Substituting (\ref{spinoransatz})
into the Killing spinor equation (\ref{Killingspinoreqn}) leads to the following algebraic and differential equations for the spinor field $\chi$ on $Y_7$:
\bea\label{spinoreqns}
\frac{1}{2}\gamma^n\partial_n\Delta \chi -\frac{\ii m}{6}\ex^{-3\Delta}\chi+\frac{1}{288}\ex^{-3\Delta}
F_{npqr}\gamma^{npqr}\chi+\chi^c&=&0~,\nn\\
\nabla_m\chi +\frac{\ii m}{4}\ex^{-3\Delta}\gamma_m\chi -\frac{1}{24}\ex^{-3\Delta} F_{mpqr}\gamma^{pqr}\chi 
-\gamma_m\chi^c&=&0~.
\eea
For a supergravity solution one must also solve the equations of motion resulting from (\ref{action}), 
as well as the Bianchi identity $\diff G=0$. 

Motivated by the discussion in the introduction, in this paper we will focus on $\mathcal{N}=2$ supersymmetric AdS$_4$ solutions for which there are 
two independent solutions $\chi_1$, $\chi_2$ to (\ref{spinoreqns}). In particular, the general ${\cal N}=2$ Killing 
spinor ansatz may be written as
\bea\label{n2spinoransatz}
\epsilon & = & \sum_{i=1,2} \psi_i^+\otimes \ex^{\Delta/2}\chi_i + (\psi^+_i)^c\otimes \ex^{\Delta/2}\chi^c_i~.
\eea
In this case 
there is a $\mathtt{u}(1)$ R-symmetry which rotates these spinors 
as a doublet. It is then convenient to introduce 
\bea\label{chi}
\chi_\pm &\equiv & \frac{1}{\sqrt{2}}\left(\chi_1\pm \ii \chi_2\right)~,
\eea
which will have charges $\pm2$ under the Abelian R-symmetry. 
For an $\mathcal{N}=2$ solution one can show that 
the spinor equations (\ref{spinoreqns}) imply that, without loss of generality, one 
can normalize $\bar\chi_\pm\chi_\pm =1$ \cite{WIP}. We shall impose this 
normalization in what follows.

\section{Contact structure}

In this section we show that any $\mathcal{N}=2$ supersymmetric AdS$_4$ solution 
with $m\neq 0$ admits a canonically defined contact structure. Moreover, the Reeb
vector field  $\xi$ for this contact structure is also a Killing vector field which 
preserves all bosonic fields, and the spinors $\chi_\pm$ in (\ref{chi}) 
have charges $\pm 2$ under $\xi$. We thus interpret $\xi$ as the dual of the 
expected $\mathtt{u}(1)$  R-symmetry. 

\subsection{R-symmetry Killing vector}

We begin by defining the one-form bilinear and its dual vector field
\bea\label{K3}
K &\equiv & \ii\, \bar\chi_+^c\gamma_{(1)}\chi_-~, \qquad \xi \ \equiv\  g^{-1}_{Y_7}(K,\cdot\, )~,
\eea
where we denote $\gamma_{(n)}\equiv \frac{1}{n!}\gamma_{m_1\ldots m_n}\diff y^{m_1}\wedge\ldots\wedge \diff y^{m_n}$.
{\it A priori} the one-form $K$ in (\ref{K3}) is complex; however, one can show that the spinor equations 
(\ref{spinoreqns}) imply that $\Imag K=0$ so that $K$ is real. It is then straightforward to 
show that $K$ is a Killing one-form for the metric $g_{Y_7}$ on $Y_7$, and hence that the dual 
vector field $\xi$ is a Killing vector field. We note for future reference the square norm
\bea\label{norm}
\|\xi\|^2 &\equiv & g_{Y_7}(\xi,\xi) \ = \ |\bar\chi_+^c\chi_+|^2 + \frac{m^2}{36}\ex^{-6\Delta}~.
\eea
In particular when $m\neq 0$ we see that $\xi$ is nowhere zero, and thus defines a one-dimensional foliation of $Y_7$.

The algebraic equation in (\ref{spinoreqns}) 
leads immediately to $\mathcal{L}_\xi \Delta=0$, and using both equations in (\ref{spinoreqns}) 
one can show that
\bea
\diff (\ex^{3\Delta}\, \bar\chi_+^c\gamma_{(2)}\chi_-) &=& - \ii \xi\lrcorner F~.
\eea
It follows that
\bea
\mathcal{L}_\xi F &=& \diff (\xi\lrcorner F)+ \xi\lrcorner\diff F \ = \ 0~,
\eea
provided the Bianchi identity $\diff F=0$ holds\footnote{In fact this is implied by supersymmetry, as we will show shortly -- {\it cf}. equation (\ref{mF}).}.  Thus $\xi$ preserves all of the bosonic fields. 
One can also show that
\bea
\mathcal{L}_{\xi} \chi_\pm &=& \pm 2\ii\, \chi_\pm~,\label{charge2}
\eea
so that $\chi_\pm$ have charges $\pm 2$ under $\xi$. We thus identify $\xi$ 
as the canonical vector field dual to the R-symmetry of the $\mathcal{N}=2$ SCFT.

\subsection{Contact form}

Provided $m\neq 0$ we may define the real one-form bilinear
\bea
\sigma &\equiv & -\frac{6}{m}\ex^{3\Delta}\, \bar\chi_+\gamma_{(1)}\chi_+~.
\eea
Using the spinor equations one can readily show that
\bea
\diff \sigma &=& -\frac{12}{m}\ex^{3\Delta}\, \Real \bar\chi_+^c\gamma_{(2)}\chi_-~,
\eea
and an algebraic computation then leads to
\bea\label{contactvolume}
\sigma\wedge (\diff\sigma)^3 &=& \frac{2^73^4}{m^3}\ex^{9\Delta}\vol_7~,
\eea
where $\vol_7$ denotes the Riemannian volume form of $Y_7$. 
It follows that $\sigma\wedge (\diff\sigma)^3$ is a nowhere-zero 
top degree form on $Y_7$, and thus by definition $\sigma$ is a contact form on $Y_7$.
Again, straightforward algebraic computations lead to 
\bea
\xi\lrcorner\sigma \ = \ 1~, \qquad  \xi\lrcorner\diff\sigma \ = \ 0~.
\eea
This implies that the Killing vector field $\xi$ is also the unique Reeb vector field 
for the contact structure defined by $\sigma$.

\section{Free energy on $S^3$}

In this section we present a general supergravity formula for the free energy $\mathcal{F}$ 
of the dual $\mathcal{N}=2$ SCFT on $S^3$. When $m\neq 0$, which 
is equivalent to a non-zero M2-brane charge of the AdS$_4$ background, 
this may be expressed in terms of the contact volume $
\int_{Y_7} \sigma\wedge (\diff\sigma)^3$ via (\ref{freeenergy}).

\subsection{Newton constant}

The effective four-dimensional Newton constant $G_4$ is computed by dimensional 
reduction of eleven-dimensional supergravity on $Y_7$. More precisely, by definition $1/16\pi G_4$ is 
the coefficient of the four-dimensional Einstein-Hilbert term, in Einstein frame. A standard
computation determines this to be
\bea\label{newton}
\frac{1}{16\pi G_4} &=& \frac{\pi\int_{Y_7} \ex^{9\Delta}\vol_7}{2(2\pi\ell_p)^9}~,
\eea
where recall that $\ell_p$ denotes the eleven-dimensional Planck length. 

On the other hand, via the AdS/CFT correspondence $G_4$ also determines the free energy $\mathcal{F}$ of the 
dual CFT on $S^3$:
\bea\label{F}
\mathcal{F}&\equiv & - \log |Z| \ = \ \frac{\pi}{2G_4}~.
\eea
More precisely, the left hand side of (\ref{F}) is minus the free energy of the unit radius AdS$_4$ computed in Euclidean quantum gravity, 
where $Z$ is the gravitational partition function. The latter is regularized to give the finite result on 
the right hand side of (\ref{F}) using the boundary counterterm subtraction method of \cite{Emparan:1999pm}. 
Combining (\ref{newton}) and (\ref{F}) leads to the supergravity formula
\bea\label{merlin}
\mathcal{F}&=& \frac{4\pi^3\int_{Y_7} \ex^{9\Delta}\vol_7}{(2\pi\ell_p)^9}~.
\eea

\subsection{Flux quantization}

Using the spinor equations (\ref{spinoreqns}) one can derive the general expression
\bea\label{mF}
m F &=& 6\, \diff (\ex^{6\Delta}\Imag \bar\chi_+^c\gamma_{(3)}\chi_-)~.
\eea
Thus provided $m\neq 0$ we see that $F$ automatically obeys the Bianchi identity 
$\diff F=0$, and moreover $F$ is in fact exact. There is thus no Dirac quantization 
condition for the four-form $F$ when $m\neq 0$.\footnote{This is certainly
not the case for solutions with $m=0$, as discussed in section \ref{discussion}.} 

On the other hand, the total M2-brane charge of the AdS$_4$ background is
\bea\label{Ngeneral}
N&=& - \frac{1}{(2\pi\ell_p)^6}\int_{Y_7} *_{11}G+\frac{1}{2}C\wedge G~.
\eea
Dirac quantization requires this to be an integer. Equation (\ref{mF}) implies 
that $F=\diff A$ where we may take the three-form potential $A$ to be  the globally defined form
\bea\label{A}
A &\equiv & \frac{6}{m}\ex^{6\Delta} \Imag \bar\chi_+^c\gamma_{(3)}\chi_-~.
\eea
Note that using (\ref{charge2}) it immediately follows that this choice of gauge is invariant 
under $\xi$, that is,
\bea
{\cal L}_\xi A & = & 0~.
\eea
Of course, one is free to add to $A$ any closed three-form $c$, which will result in the same 
curvature $F$:
\bea\label{Atrans}
A &\rightarrow & A + \frac{1}{(2\pi \ell_p)^3}c~.
\eea
If $c$ is exact this is a gauge transformation of $A$ and leads to a physically equivalent 
M-theory background. In fact more generally if $c$ has integer periods then the 
transformation (\ref{Atrans}) is a large gauge transformation of $A$, again leading 
to an equivalent solution. It follows that only the cohomology class of $c$ in the torus
$H^3(Y_7;\R)/H^3(Y_7;\Z)$ is a physically meaningful parameter, and this corresponds 
to a marginal parameter in the dual CFT. In fact the free energy will be independent of this 
choice of $c$, which is why we have set $c=0$ in (\ref{A}). There is also the possibility 
of adding discrete torsion to $A$ when $H^4_{\mathrm{torsion}}(Y_7;\Z)$ is non-trivial, but we will 
not discuss this here.

Substituting our ansatz (\ref{ansatz})  into the general expression (\ref{Ngeneral}) leads to
\bea
N&=& \frac{1}{(2\pi \ell_p)^6}\int_{Y_7} m\ex^{3\Delta}\vol_7 - \frac{1}{2}A\wedge F~.
\eea
Using (\ref{A}) and the algebraic equation in (\ref{spinoreqns}) one can easily compute
\bea\label{Nexplicit}
N &=& \frac{1}{(2\pi \ell_p)^6}\frac{m^2}{2^53^2}\int_{Y_7} \sigma\wedge (\diff\sigma)^3~.
\eea
Combining (\ref{Nexplicit}), (\ref{merlin}) and (\ref{contactvolume}) now leads
straightforwardly to (\ref{freeenergy}).

\section{Scaling dimensions of BPS M5-branes}

A probe M5-brane whose world-space is wrapped on a generalized calibrated five-submanifold $\Sigma_5 \subset Y_7$ and which moves along a geodesic
 in AdS$_4$ is expected to correspond to 
a
BPS operator  $\mathcal{O}_{\Sigma_5}$
in the dual three-dimensional SCFT.  
In particular, when $Y_7$ is a Sasaki-Einstein manifold, the scaling dimension of this operator can be calculated from the volume of the 
five-submanifold $\Sigma_5$ \cite{Berenstein:2002ke}. In this section 
we show that a simple generalization of this correspondence holds for the general ${\cal N}=2$ supersymmetric 
AdS$_4\times Y_7$ 
solutions\footnote{Such supersymmetric M5-branes exist only for certain boundary conditions 
\cite{Klebanov:2010tj,Benishti:2010jn}, and  our discussion
 here applies to these cases.} treated in this paper, and in particular we prove the formula (\ref{Delta}). 
The calculation is a simple adaptation of that presented in
\cite{Martelli:2003ki}, and more details will appear in \cite{WIP}.   

\subsection{Generalized calibration}

Given a Killing spinor $\epsilon$ of eleven-dimensional supergravity, it is simple to derive the BPS bound  \cite{Barwald:1999ux,Martelli:2003ki} 
\bea
\epsilon^\dagger  \epsilon \, L_\text{DBI}\, \vol_5 &  \geq  &  \frac{1}{2} (\hat{k}\lrcorner  H)\wedge H +  \hat{\mu} \wedge H   +   \hat{\nu} ~. \label{mainbound}
\eea
This bound is saturated if and only if $\proj \epsilon = 0$, where $\proj \equiv(1 - \tilde \Gamma)/2$ is the $\kappa$-symmetry projector,  
and corresponds to a probe M5-brane preserving supersymmetry.
Here $H$ is the three-form on the M5-brane, defined by $H= h + j^*C$ where $h$ is closed and $j^*$ denotes the pull-back to the M5-brane world-volume. 
The one-form $\hat{k}$, two-form $\hat{\mu}$ and  five-form $\hat{\nu}$ denote the \emph{pull-back} to 
$\Sigma_5$ of the differential forms \cite{Gauntlett:2002fz} defined by the bilinears  $k =   \bar \epsilon \Gamma_{(1)}\epsilon $,
$ \mu  =    \bar \epsilon \Gamma_{(2)}\epsilon$, and  $\nu  =   \bar \epsilon \Gamma_{(5)}\epsilon$, respectively, 
and  $\vol_5$ is the volume form on the world-space of the M5-brane.
We have defined $\bar \epsilon \equiv \epsilon^\dagger \Gamma_0$ as usual.

We will use the static gauge embedding 
$\{ \tau = \sigma^0 , x^m = \sigma^m \}$,  where  $\tau$ is global
 time in AdS$_4$ and  $x^m$, with $m=1,\dots, 5$, are coordinates on $Y_7$.
  The Dirac-Born-Infeld Lagrangian is given by $L_{\text{DBI}} = \sqrt{\det(\delta_m^{\ n} + H_m^{* n})}$,  
where the two-form $H^* \equiv *_5 H$ is the world-space dual of $H$. 
Using the explicit form of the eleven-dimensional ${\cal N}=2$ Killing spinor (\ref{n2spinoransatz})
one can show that the bound (\ref{mainbound}) is saturated when $\rho=0$
({\it i.e.} the M5-brane is at the centre of AdS$_4$) and 
\bea
\frac{\me^{\Delta}}{2} L_\text{DBI}\, \vol_5 &  = &  \frac{1}{2} (\hat k\lrcorner H)\wedge H + \hat \mu \wedge H   +  \hat \nu ~.\label{satur}
\eea

\subsection{Energy of a BPS M5-brane}

The energy density of an M5-brane can be computed by solving the  Hamiltonian constraints, leading to
 \bea
 {\cal E} & =&  P_\tau \ =\ T_{\text{M5}} \left( \frac{\ex^\Delta  }{2} L_{\text{DBI}} + \mathcal{C}_\tau\right)~,
 \eea 
 where  $T_{\text{M5}}=2\pi/(2\pi \ell_p)^6$ is the M5-brane tension and the contribution from the Wess-Zumino coupling is $
{\cal C}_\tau \vol_5  = \de_\tau \lrcorner C_6 - \tfrac{1}{2} (\de_\tau \lrcorner C) \wedge (C- 2H)$,
with the potential $C_6$ defined through $\diff C_6 = *_{11}G +\tfrac{1}{2}C\wedge G$. However, from the explicit expression of
 $C$ we presented earlier one can check that we have ${\cal C}_\tau=0$.
The M5-brane energy is then given by 
\bea
E_\text{M5}& =&  T_{\text{M5}} \int_{\Sigma_5} \frac{\ex^\Delta  }{2} L_{\text{DBI}}\, \vol_5 \, =\, 
T_{\text{M5}} \int_{\Sigma_5} \frac{1}{4} (\xi \lrcorner H)\wedge H + \hat \mu \wedge H   +  \hat \nu ~,
\label{trivial}
\eea
where we used that the time-like Killing vector $k^\#$ dual to the one-form $k$ is given by
$k^\# = \partial_\tau + \tfrac{1}{2} \xi$.
Let us briefly discuss this expression for the energy.   With our gauge choice (\ref{A}) for the three-form potential, 
in general we have $H=A+h$, where $h$ is a closed three-form.  If $h$ is exact and 
invariant\footnote{One should obviously require that $\de /\de \tau$ and $\xi$ generate symmetries of the M5-brane action.} 
under $k^\#$, namely $h=\diff b$ with ${\cal L}_{k^\#} b=0$, then one can check that the integral does not depend on $h$. To see this, 
one has to recall that ${\cal L}_{k^\#} A=0$, use the results of
\cite{Gauntlett:2002fz}, and  apply Stokes' theorem repeatedly. If $h$ is not exact, \emph{a priori} it will contribute to the energy, 
and hence we expect the dimension of the dual operator to be affected. We  leave an investigation of this interesting possibility for future work, 
and henceforth set $H=A$. 

After some straightforward computations
\cite{WIP} the integrand in (\ref{trivial}) can be evaluated in terms of the contact structure, 
and we get the remarkably simple result\footnote{The sign arises from our choice of conventions, {\it cf.} \cite{Gabella:2009wu}.} 
\bea
E_\text{M5}  &=&  - T_{\text{M5}}\frac{m^2}{2^6 3^2} \int_{\Sigma_5} \sigma \wedge (\diff \sigma )^2 ~.
\eea
Combining the latter with (\ref{Nexplicit}), and using the AdS/CFT dictionary 
$\Delta(\mathcal{O}_{\Sigma_5})=E_\text{M5}$,
leads straightforwardly to the formula (\ref{Delta}) for the scaling dimension.

\section{Applications}

As in \cite{Gabella:2009ni}, the formulae (\ref{freeenergy}) and (\ref{Delta}) have some immediate 
applications. 

\subsection{Topological and localization formulae}

Let us suppose that the Reeb vector field $\xi$ is quasi-regular, which means that 
all its orbits are closed and hence $\xi$ integrates to a $U(1)=U(1)_R$ isometry of $Y_7$. 
Since (\ref{norm}) implies that $\xi$ is nowhere zero, it follows that in this case $Y_7$ is the total space 
of a $U(1)$ principal orbifold bundle $\mathcal{L}$ over a six-dimensional orbifold $V\equiv Y_7/U(1)_R$; the latter is smooth 
precisely when $U(1)_R$ acts freely on $Y_7$. If we denote by $v$ the canonically normalized generator 
of the $U(1)_R$ action, so that we may write $v=\partial/\partial\varphi$ where 
the coordinate $\varphi$ has period $2\pi$, then $\xi=kv$ for some constant $k>0$, and 
the contact volume may be written
\bea
\frac{k^4}{(2\pi)^4}
\int_{Y_7} \sigma\wedge (\diff\sigma)^3 &=& \int_V c_1(\mathcal{L})^3\ \in\, \mathbb{Q}~.
\eea
Here we have used the general fact that the first Chern class $c_1(\mathcal{L})$ of a principal 
$U(1)$ orbifold bundle $\mathcal{L}$ over an orbifold $V$ is a rational cohomology class. 
The constant $k$ must also be rational, since one computes
\bea
\mathcal{L}_\xi \left( \bar\chi_\pm^c\chi_\pm\right) &= & \pm 4\ii\, \bar\chi_\pm^c\chi_\pm~,\eea
which implies that $\bar\chi_\pm^c\chi_\pm$ has charge $\pm4$ under $U(1)_R$. 
On the other hand, $\bar\chi_\pm^c\chi_\pm$  must have an integer charge under 
$v$, in order to be single-valued in $\varphi$, implying that $4/k\in \mathbb{N}$.

The upshot is that for gravity solutions with a $U(1)_R$ isometry, 
the coefficient of $\pi N^{3/2}$ in the free energy (\ref{freeenergy}) is the square 
root of a rational number, and that the latter has a topological interpretation as a Chern number. 
The corresponding result for supersymmetric AdS$_5$ solutions of Type IIB string theory in 
\cite{Gabella:2009ni} is that the central charge $a$ computed via supergravity is 
rational when one has a $U(1)_R$ isometry. In the dual $d=4$, $\mathcal{N}=1$ SCFT 
this is clear, since there is a well-known cubic expression for $a$ in terms 
of R-charges with rational coefficients \cite{Anselmi:1997ys}. On the other hand, 
it is currently unclear, at least to the authors, why the coefficient of $\pi N^{3/2}$
in the free energy
should be the square root of a rational number when one has 
rational R-charges in the $d=3$, $\mathcal{N}=2$ SCFT. 
We may thus regard this as a prediction of supergravity for field theory.

Also as in \cite{Gabella:2009ni}, we may write the contact volume in terms of a
Duistermaat-Heckman integral on the cone $X$ over $Y_7$
\bea\label{DH}
\int_{Y_7} \sigma\wedge (\diff\sigma)^3 &=& \int_X\ex^{-r^2/2}\, \frac{\omega^4}{4!}~.
\eea
Here $r>0$ is a coordinate on $\R_+$ in $X\cong \R_+\times Y_7$,
 $\omega=\frac{1}{2}\diff(r^2\sigma)$ is a symplectic 
form on $X$, and $r^2/2$ is a Hamiltonian function 
for the Reeb vector field $\xi$. The right hand side 
of (\ref{DH}) may then often be computed via localization. 
Roughly, this involves choosing an equivariant 
symplectic resolution of the singularity of $X$ at $r=0$. We refer to 
\cite{Gabella:2009ni} and references therein for a more detailed 
discussion, especially in the case that $X$ is symplectic toric. 
In practice, this is often a very useful method for computing 
the left hand side of (\ref{DH}) using only topological methods. 

\subsection{Massive deformations of CY$_3\times \C$}

As a concrete example, in this section we briefly 
consider the supergravity solutions of \cite{Corrado:2001nv}. The original solution 
in this paper is a warped AdS$_4\times \tilde S^7$ background with internal $G$-flux on the 
(squashed and stretched) seven-sphere
$\tilde S^7$. This has a dual field theory interpretation as deforming 
the ABJM theory, dual to the round $S^7$ solution, by a mass deformation and flowing to the IR, with 
the  warped AdS$_4$ solution of \cite{Corrado:2001nv} describing the IR fixed point
 \cite{Klebanov:2008vq}. 
In fact more generally one can consider M2-branes probing the Calabi-Yau four-fold geometry CY$_3\times \C$, where 
CY$_3$ denotes \emph{any} Calabi-Yau three-fold cone. One expects these to have dual field theory descriptions 
in which one can give a mass to a gauge-invariant scalar chiral primary operator, the latter being 
dual to a Kaluza-Klein mode arising from the holomorphic function $z_0$ on $\C$. This will trigger a 
renormalization group flow, whose end-point has a gravity dual described by 
a generalization of the warped $\tilde S^7$ solution of  \cite{Corrado:2001nv}. The latter is in fact 
then the special case CY$_3=\C^3$.
Other special cases of such solutions, and their field theory duals, have been discussed 
recently in \cite{Ahn:2009bq,Jafferis:2011zi}. We shall 
discuss the general case in more detail in \cite{WIP}. 

Here we prove 
that these renormalization group flows are universal in the sense that the 
ratio of the free energies in the IR and UV is independent of the choice of three-fold CY$_3$. 
This was anticipated recently in \cite{Jafferis:2011zi}. A key point is that we do not need 
the generalization of the explicit supergravity solution in \cite{Corrado:2001nv}, but rather 
the universal formula
\bea\label{ratio}
\frac{\mathcal{F}_{\mathrm{IR}}}{\mathcal{F}_{\mathrm{UV}}} &=& \sqrt{\frac{16}{27}}
\eea
in fact follows straightforwardly from the contact volume formula (\ref{freeenergy}).

To see this, consider a general CY$_3\times \C$ Calabi-Yau four-fold, whose (generically singular) 
Sasaki-Einstein link $Y_7^\mathrm{SE}$ describes the UV background. This has at least a $\C^*\times\C^*$ symmetry,
in which the first $\C^*$ acts on the CY$_3$, and under which the CY$_3$ Killing spinors have charge $\tfrac{1}{2}$, 
and the second $\C^*$ acts in the obvious way on the copy of $\C$. Let us denote the 
components of the Reeb vector field for the Calabi-Yau four-fold in this basis as $(\xi_1,\xi_0)$. 
It is straightforward to see that the CY$_4$ Killing spinors have charge $2$, as in equation (\ref{charge2}), 
precisely when
\bea\label{weight4}
\xi_1+\xi_0 &=& 4~,
\eea
which is also equivalent to the holomorphic $(4,0)$-form $\Omega_{(4,0)}=\Omega_{(3,0)}\wedge \diff z_0$ having charge 4. 
As shown in appendix B of \cite{Gabella:2010cy}, in general the contact volume is a function of the Reeb vector field.
In our case the contact volume of $Y_7$ is given by the general formula
\bea\label{volumes}
\mathrm{Vol}(Y_7)[\xi_1,\xi_0] &=& \frac{1}{\xi_0}\, \mathrm{Vol}(Y_5)[\xi_1]~,
\eea
where $Y_5$ denotes the five-manifold link of CY$_3$. 
Then one easily shows that 
$\xi_1=3$ for a Sasaki-Einstein metric, so that (\ref{volumes}) implies the relation $\mathrm{Vol}(Y_7^{\mathrm{SE}})= \mathrm{Vol}(Y_5^{\mathrm{SE}})$ between Sasaki-Einstein volumes. Notice here that $\xi_0=1$ 
follows from (\ref{weight4}), and that this indeed gives the expected scaling dimension $\Delta=\tfrac{1}{2}$ of a free chiral field.\footnote{Note 
there is a factor of $\tfrac{1}{2}$ in going from the geometric scaling dimension under $r\partial_r$ to the scaling dimension $\Delta$ in field theory, 
{\it cf.} equation (2.31) of \cite{Gauntlett:2006vf}.}

Let us now consider the IR solution corresponding to the mass deformation. Since the operator dual to 
$z_0$ is given a mass, the scaling dimension necessarily changes from $\Delta=\tfrac{1}{2}$ to $\Delta=1$. 
Thus one expects $\xi_0=2$ in the supergravity solution, and one indeed sees that this is the case. 
The Killing spinors always have charge $2$ (\ref{charge2}); thus equation (\ref{weight4}) still holds, and 
we conclude that $\xi_1=2$ for the mass-deformed background. Using this, together with the fact 
that the contact volume of $Y_5$ has homogeneous degree $-3$ \cite{Gabella:2010cy} under $\xi_1$, we conclude from (\ref{volumes}) 
that
\bea
\mathrm{Vol}(Y_7^{\mathrm{mass}}) &=& \frac{1}{2}\, \mathrm{Vol}(Y_5)[2] \ = \ \frac{1}{2}\cdot\left(\frac{2}{3}\right)^{-3}\mathrm{Vol}(Y_5)[3] \ = \ 
\frac{27}{16}\, \mathrm{Vol}(Y_7^{\mathrm{SE}})~.
\eea
Using (\ref{freeenergy}) then leads directly to (\ref{ratio}).

\section{Discussion}
\label{discussion}

In this paper we have taken a first glimpse at the geometry characterizing general ${\cal N}=2$ supersymmetric
AdS$_4$ solutions
of M-theory with non-zero M2-brane charge. In particular, we have shown that these admit a Killing 
vector $\xi$ that realizes the  $\mathtt{u}(1)$ 
 R-symmetry of the dual field theories, and a contact one-form 
$\sigma$, in terms of which the free energy and the scaling dimensions of certain BPS operators 
can be written. The geometry of $Y_7$ is a natural generalization of Sasaki-Einstein 
geometry, precisely as was found in \cite{Gabella:2009ni,Gabella:2009wu}
 for general supersymmetric AdS$_5\times Y_5$ solutions of type IIB  supergravity. 
 As an application of our results we have briefly discussed a class of 
solutions discovered in \cite{Corrado:2001nv}. An investigation of new solutions is currently under way \cite{WIP}.
We should point out that there exist in the literature  ${\cal N}=2$ supersymmetric AdS$_4\times Y_7$ backgrounds
where the M2-brane charge vanishes. An example is
the solution discussed in  \cite{Gauntlett:2000ng} (originally 
found in \cite{Pernici:1984nw}), representing the near-horizon limit of M5-branes wrapped on a special Lagrangian submanifold inside 
a Calabi-Yau three-fold times $\R^2$. 
Our results do not apply to this solution, and in particular we expect that the free energy in this case will scale as $N^3$.
However we leave the study of this class of solutions for the future. 

In \cite{WIP} the geometry of $Y_7$ will be explored in greater detail. In particular, we will show
that this geometry is characterized by a local $SU(2)$ structure, which turns out to be strikingly similar to the 
$SU(2)$ structure characterizing supersymmetric AdS$_5\times Y_6$ solutions of M-theory \cite{Gauntlett:2004zh}.
From the findings of this paper, and of  \cite{WIP}, arise a number of interesting questions to address. 
In \cite{Gabella:2009wu} the results of \cite{Gauntlett:2005ww} and \cite{Gabella:2009ni} have been elegantly reformulated in terms of 
generalized geometry of the cone $C(Y_5)$, where the metric and NS $B$-field are unified.
Similarly, it is tempting to speculate that the geometry of the eight-dimenional cones $C(Y_7)$ will turn 
out to be an analogous kind  of generalized geometry, that treats the metric and $C$-field 
on equal footing
\cite{West:2001as,Hull:2007zu,Pacheco:2008ps,Berman:2010is}.  We also anticipate a generalization of volume minimization of Sasaki-Einstein 
manifolds \cite{Martelli:2005tp,Martelli:2006yb}, along the lines of 
\cite{Gabella:2010cy}. Finally, it would be very interesting to understand whether a direct relationship 
between  localization  on the field theory side 
\cite{Kapustin:2009kz} and  localization on the gravity side, discussed here
 and in \cite{Martelli:2005tp,Martelli:2006yb,Lee:2006ys,Gabella:2009ni},  can be established. 

\subsection*{Acknowledgments}
\noindent
We thank Nessi Benishti for collaboration at an early stage of this work. 
D.~M.  thanks the Galileo Galilei Institute for Theoretical Physics for  hospitality
 and the INFN for partial support during the initial stages of this work.
M.~G., D.~M., and J.~F.~S. would like to thank  the
Centro de ciencias de Benasque Pedro Pascual
 for hospitality during the completion of this 
work. M.~G. is supported by the Swiss National Science Foundation, D.~M. by EPSRC Advanced
 Fellowship EP/D07150X/3, and J.~F.~S. by a Royal Society University Research Fellowship.

\end{document}